\documentclass[sigconf]{acmart}
\AtBeginDocument{%
  \providecommand\BibTeX{{%
    \normalfont B\kern-0.5em{\scshape i\kern-0.25em b}\kern-0.8em\TeX}}}

\copyrightyear{2021} 
\acmYear{2021} 
\setcopyright{acmcopyright}\acmConference[PEARC '21]{Practice and Experience in Advanced Research Computing}{July 19--22, 2021}{Virtual Conference}
\acmBooktitle{Practice and Experience in Advanced Research Computing (PEARC '21), July 19--22, 2021, Virtual Conference}
\acmPrice{15.00}
\acmDOI{10.1145/3311790.3400848}
\acmISBN{978-1-4503-6689-2/20/07}

\begin{document}

\title{Experiences with managing data parallel computational workflows for High-throughput Fragment Molecular Orbital (FMO) Calculations}
\author{Dimuthu Wannipurage}
\affiliation{%
  \institution{Indiana University}
  \country{USA}}
\email{dwannipu@iu.edu}

\author{Indrajit Deb}
\affiliation{%
 \institution{University of Michigan}
 \country{USA}}
\email{ideb@umich.edu}

\author{Eroma Abeysinghe}
\affiliation{%
  \institution{Indiana University}
  \country{USA}}
\email{eabeysin@iu.edu}

\author{Sudhakar Pamidighantam}
\affiliation{%
  \institution{Indiana University}
  \city{Bloomington}
  \state{Indiana}
  \country{USA}}
\email{pamidigs@iu.edu}

\author{Suresh Marru}
\affiliation{%
  \institution{Indiana University}
  \city{Bloomington}
  \state{Indiana}
  \country{USA}}
\email{smarru@iu.edu}

\author{Marlon Pierce}
\affiliation{%
  \institution{Indiana University}
  \country{USA}}
\email{marpierc@iu.edu}

\author{Aaron T. Frank}
\affiliation{%
 \institution{University of Michigan}
 \country{USA}}
\email{afrankz@umich.edu}

\renewcommand{\shortauthors}{Wannipurage and Deb, et al.}
\renewcommand{\shorttitle}{Fragment Molecular Orbital FMO Framework}

\begin{abstract}
Fragment Molecular Orbital (FMO) calculations provide a framework to speed up quantum mechanical calculations and so can be used to explore structure-energy relationships in large and complex biomolecular systems. These calculations are still onerous, especially when applied to large sets of molecules. Therefore, cyberinfrastructure that provides mechanisms and user interfaces that manage job submissions, failed job resubmissions, data retrieval, and data storage for these calculations are needed. Motivated by the need to rapidly identify drugs that are likely to bind to targets implicated in SARS-CoV-2, the virus that causes COVID-19, we developed a static parameter sweeping framework with Apache Airavata middleware to apply to complexes formed between SARS-CoV-2 M\textsuperscript{pro} (the main protease in SARS-CoV-2) and  2820 small-molecules in a drug-repurposing library. Here we describe the implementation of our framework for managing the executions of the high-throughput FMO calculations. The approach is general and so should find utility in large-scale FMO calculations on biomolecular systems.
\end{abstract}



\maketitle

\section{Introduction}
SARS-CoV-2 has infected more than 142 million people worldwide and has killed more than 3 million individuals as of April 2021. Immediately, the repurposing of old drugs offers us our best chances to reduce the severity of COVID-19 in patients who have already been infected with SARS-CoV-2.


\begin{figure*}[htbp]
 \centering
 \includegraphics[width=\textwidth]{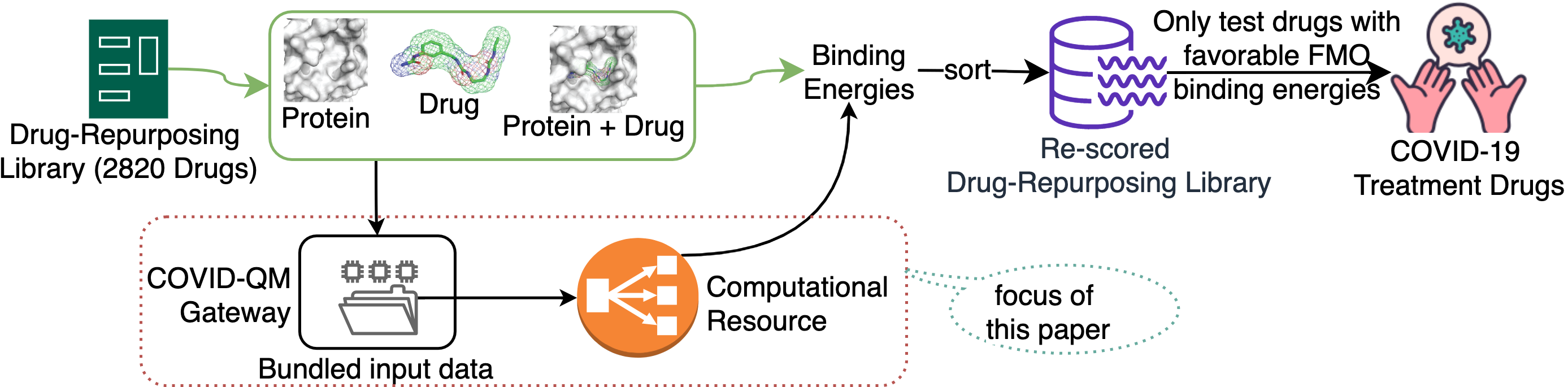}
 \caption{Our larger vision is to create a scalable Framework for High-throughput Fragment Molecular Orbital (FMO) Calculations. In this paper we focus on the data parallel execution aspects.}
 \label{figBigPicture}
\end{figure*}

\textbf {Scientific goal:} We use high-level quantum calculations to guide drug repurposing efforts for COVID-19. 
The rationale here is to predict which known drugs are most likely to bind to known COVID-19 molecular targets. Such drugs are, in turn, expected to inhibit the replication of SARS-CoV-2 and, in so doing, stop the spread of COVID-19 \cite{baker2020repurposing, bobrowskia2020computational,filippo2020silico,duarte2020repurposing,smith2020repurposing}. However, molecular docking methods rely on inaccurate scoring functions that frequently lead to false-positive and, presumably, false-negative predictions. As an alternative to the simple empirical scoring functions, quantum mechanical methods that take into account higher-order quantum mechanical (QM) effects can be employed. The more accurate QM-derived estimates of drug binding energies can then be used to re-score and re-prioritize compounds in a given library based on their docked poses. 

Until recently, however, such calculations were prohibitive due to computational cost. Methodological advances, such as divide-and-conquer techniques, now make it feasible to calculate drug-binding energies using QM. In particular, the fragment-molecular-orbital (FMO) method has shown tremendous promise as a drug screening tool that outperforms approaches relying on approximate binding energy estimates \cite{fedorov2004second,okimoto2018use,heifetz2020guiding}. We, therefore, set out to use the FMO method to estimate the binding energies of compounds in a drug repurposing database called SuperDRUG2.0 that contains FDA-approved and marketed drugs \cite{siramshetty2018superdrug2,goede2005superdrug,cheminfo001}. A library of 3993 small-molecule compounds for which 3D structures were available were filtered to choose small molecules having molecular weight ranging from 200 g/mol to 500 g/mol and atomic numbers more than 10. The library was further filtered to remove the small-molecules forming salt and in complex with heavy elements. Finally, a set of 2820 drug-like small molecules was selected and docked to the active site of SARS-CoV-2 M\textsuperscript{pro} following standard docking protocol\cite{ruiz2014rdock}, and the resulting protein-ligand complexes were subjected to the FMO calculations. This is illustrated as Drug-Repurposing libary in Fig. \ref{figBigPicture}. Based on the estimates from FMO calculations, the drugs that are likely to bind to known molecular targets in SARS-CoV-2 can be selected and advanced as strong candidates for clinical trials.

\textbf{Computing challenge:} 
The project required the execution of 2820 $\times$ 2 = 5640 independent jobs using the FMO implementation included in the General Atomic and Molecular Electronic Structure System (GAMESS) \cite{GAMESS} an initio quantum chemistry software package. Half of the jobs, corresponding to the protein-ligand complexes, required 48 processing cores, each with an execution time of $\sim$7 hrs. The other half, corresponding to the ligand alone, also used 48 processing cores, each with an execution time of $\sim$5 mins. Job monitoring was required to detect convergence failures, which occur periodically in FMO calculations. Calculation outputs had to be parsed to extract energetic and atomic charge data from the GAMESS/FMO log file for downstream analyses.


By the nature of the execution where the same application is being invoked with different input configurations, these calculations can be efficiently performed using a static parameter sweeping technique. In this particular research use case, input parameters are the various poses of all the above mentioned FDA approved drug molecules in Cartesian coordinate format, which are written into text files. These text files become the input parameters for the GAMESS applications deployed in computational resources to perform FMO calculations. In this paper we discuss our experiences with developing a framework (Fig. \ref{figBigPicture}) to submit jobs to computational resources that do not have an out-of-the-box capability to handle data-parallel parameter sweeping, we need to have an external job management engine that controls the parameter sweeping logic. In addition, the framework to manage job executions must be able to provide a user-friendly gateway interface to submit jobs and monitor job status, manage job failures and possible re-submissions, and handle large number of job and post processing final outputs. The framework must also be generic so that it can be reused for similar and related projects.




\section{JOB ORCHESTRATION}

We based our implementation on Apache Airavata \cite{marru2011apache}, which provides both a highly flexible job management framework and a turnkey user interface environment that can be used for managing job executions and data. 

\begin{figure}[h]
  \centering
  \includegraphics[width=\linewidth]{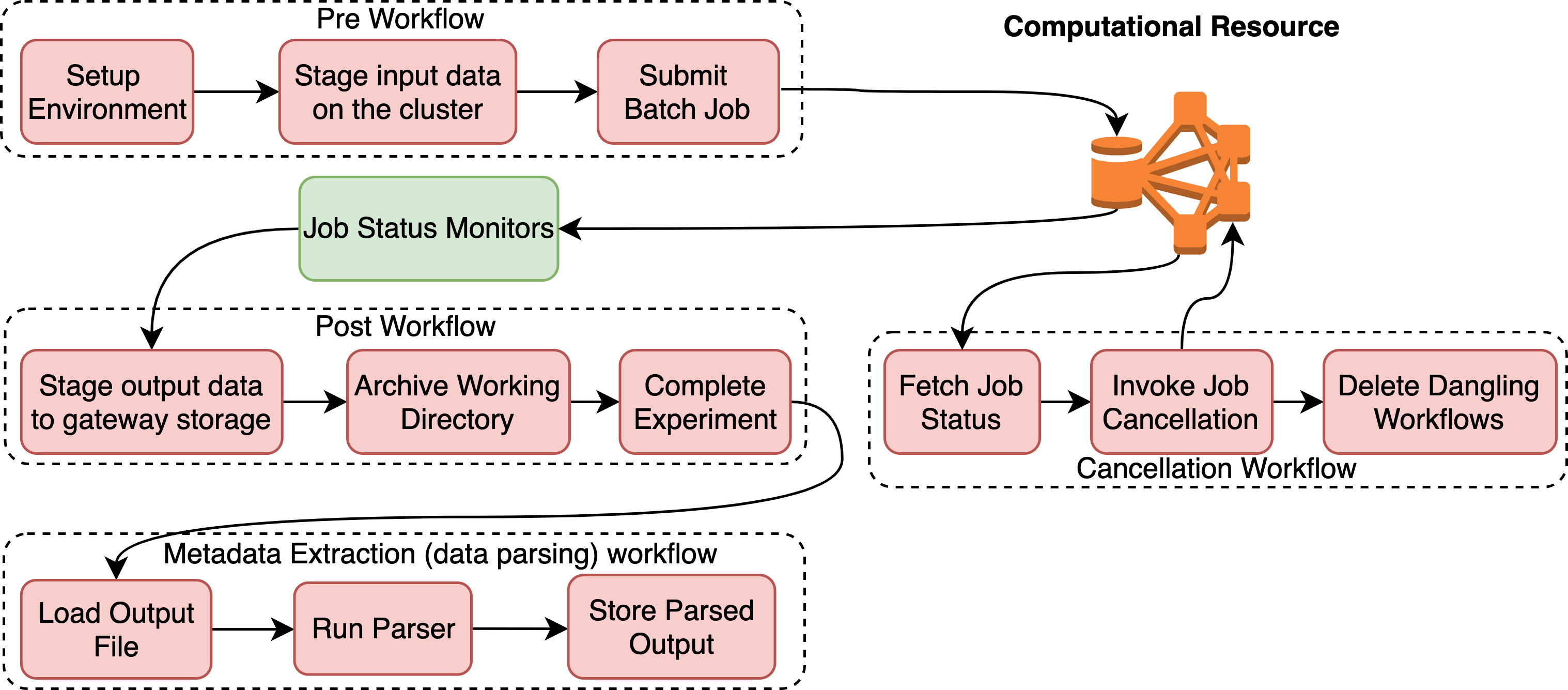}
  \caption{Workflow composition of the Job Orchestration Engine of Apache Airavata}
  \label{fig:workflows}
\end{figure}

Apache Airavata's job orchestration framework \cite{airavata-helix} is the component that handles the life cycles of the jobs submitted to High Performance Computing (HPC) clusters.  It consists of multiple workflows (Figure \ref{fig:workflows}) integrated together to handle the different stages of the life cycle of a job submitted into a remote computational resource. These job orchestration workflows are identified as pre-workflows, post-processing workflows, cancellation workflows and data-parsing workflows. The pre-workflow is responsible for executing the steps from the job initialization to the job submission on the computational resource. This includes creating the working environment on the computational resource for the job, generating the job script, transferring input files along with the job script and finally submitting the job to the computational resource. 
The post-processing workflow is responsible for staging the output files from the execution host once the job is completed. A cancellation workflow is executed when user wants to cancel an already-running job prematurely. A data-parsing workflow runs user defined data parsers on the output files generated from the job. Typically, this workflow is executed immediately after the post workflow is completed. There might be zero or more data-parsing workflows running for a single job depending on the user or gateway preferences.

\textbf{Supporting Static Parameter Sweeping at Job Orchestration}: The above job orchestration framework was designed initially assuming that an experiment contains a single job. When it comes to static parameter sweeping, there might be many similar jobs spawned from a single experiment with having changes only to input parameters. To address this issue, we analyzed how each workflow in the job orchestrator should be updated to support the required scalability at the job level of an experiment.

\textbf{Scaling jobs at the pre-workflow stage}: In this particular use case, we have to submit \textit{n} jobs in parallel to the target computational cluster with different input parameters. The pre-workflow  submits single parent job to the cluster, and that job contains the information on how to spin up child jobs that perform parameter sweeping. The environment setup task contains a batch of instructions to create the working environments for all those child jobs, which are executed through a single command in the computational resource. From the input data side, all the inputs for the parameter sweep are sent as a tar archive to the remote computational resource, where they are unarchived and moved to the working environments that were created in the environment setup phase for each child job. 



\textbf{Application and execution environment}: Jobs were run on the Stampede2 supercomputer. We used GAMESS version 2019 R02 and executed using the SKX-normal partition to utilize the Skylake Intel Xeon Platinum 8160 based nodes with 48 cores. The GAMESS default executable script, rungms, has been adapted to perform high throughput parallel processing for multiple job runs, while each run is assigned a single node. A local scratch directory in Stampede2 is created that is node-specific for each run. Each run used 24 processor cores and 24 data servers for processing using MPI parallelization of GAMESS executable. When a high throughput experiment is submitted in the COVID-QM gateway in the form of a job array, the nodelist is obtained dynamically from the SLURM environment variable \$SLURM\_JOB\_NODELIST and used to define the GAMESS execution host GMS-HOST and a node specific scratch for each of the job inputs of the job array. Each production job submission contained a job array of 120 jobs and used the maximum available nodes of 120 per submission following the reservation allocated for this COVID-19 consortium project.

Our initial approach to scale jobs at the cluster level is to utilize SLURM job arrays but in Stampede2 where we ran most of the calculations, standard SLURM job array support was disabled. As a solution, we considered using a launcher tool that is embedded with the GAMESS application installed in the cluster with a MPI backend to submit parallel jobs. In this case, the main job script executed GAMESS child jobs in a loop which runs from 0 to number of sweep jobs -1 by utilizing the MPI-backed launcher tool. Here, the loop index is used as the child job index and rest of the configurations were the same as the proposed job array based solution.

\textbf{Scaling Post Workflows}: Job monitors in the middleware detect when a job completes, and a post-workflow is executed to pull the generated outputs from the remote resource into the gateway storage. In a multi-job submission situation like this, there could be multiple post-workflows running per single experiment to fetch the outputs of each child job. As the exact job completion time can not be predicted accurately, post-workflows for these jobs are triggered randomly. Because of that, there must be a mechanism to guarantee the eventual consistency of the entire job collection.

In proposed MPI-backed launcher submissions, SLURM only sends email notifications when the parent job is completed or fails. If the parent job comes into either of those states, we can safely assume that all the child jobs also have completed or failed. However, there is no mechanism in SLURM to send notifications for the status of child jobs. As a solution for this, we injected a curl command into the beginning and end of the job script with the child job id as a path parameter. When the job is executed, these curl commands are also executed. These curl commands point to an HTTP endpoint we deployed. Using the curl messages received at the HTTP endpoint, we can detect whether the child job completed or not. This approach enables the opportunity to fetch the outputs from child jobs before the main job finishes.

\begin{figure}[h]
  \centering
  \includegraphics[width=\linewidth]{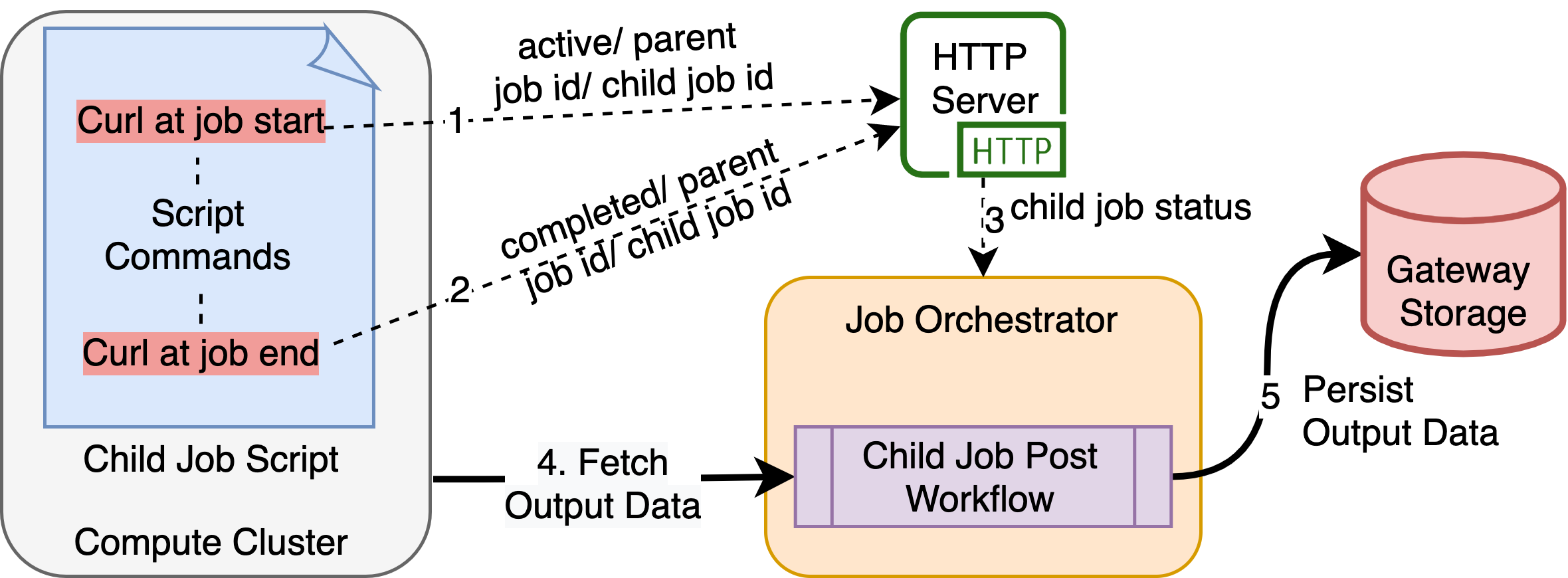}
    \setlength{\belowcaptionskip}{-8pt}
  \caption{Monitoring of child jobs using injected curl commands to the job script. Dashed lines and solid lines represent inter-service communications and data paths, respectively. Numbers on the paths represent the order of execution.}
  \label{fig:realtime-monitor}
\end{figure}

Figure~\ref{fig:realtime-monitor} depicts the curl message flow and the invocation of the respective post-workflow for the child job upon receiving the child job completion notification.

\textbf{Handling the eventual experiment completion consistency}: Even though curl-based monitoring helps us find completed child jobs for a parameter sweeping experiment, it does not give us information about failed jobs. In the case of a job crash at the computational resource level, we do not get curl messages as the job script failed before reaching that point. Even in the case of successful job execution, we can not completely guarantee that curl messages reaches the HTTP server because of possible firewall restrictions or intermittent network disruptions. For these reasons, curl-based job monitoring cannot be used as a mechanism to measure eventual completion consistency. 

\begin{figure}[h]
  \centering
  \includegraphics[width=\linewidth]{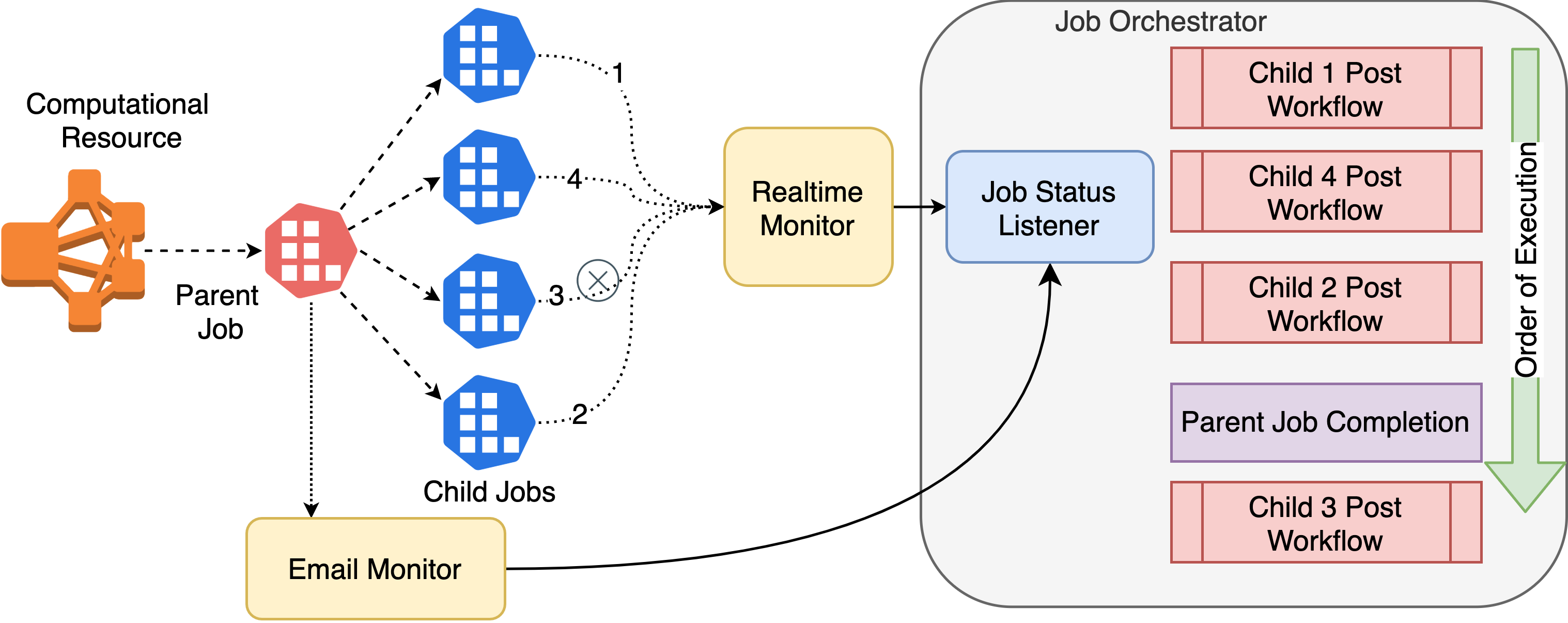}
  \caption{
  Eventual consistency model for post workflows. 
  }
  \label{fig:eventual-consistncy}
\end{figure}

Email notifications sent by the SLURM scheduler when the parent job is completed or fails can be used to achieve eventual completion consistency more reliably but with longer and inconsistent latency. The email monitor of Apache Airavata fetches this email and passes the status of the parent job into the job orchestrator to process. As the job orchestrator has information about already-processed child job post-workflows, it filters out any child job that was not processed through the curl-based monitoring. Once that subset is determined, the job orchestrator invokes the post-workflows for those remaining child jobs. When those are completed, the experiment is marked as completed or failed based on the parent job status. Figure~\ref{fig:eventual-consistncy} depicts an overview of the  consistency model described here.

\textbf{Output Data Parsing}: In some cases, the output of a GAMESS FMO calculation for certain poses of the drug molecule may not converge. When this occurs, users should be able to identify these poses and submit jobs with new input parameters that may allow convergence. We used data parsing workflows to parse output files and check whether the FMO calculations have converged or not and also to identify the reason. If the convergence issue is with the coordinates, we simply energy minimize the systems in our local computer and generate new input files in which the  calculation parameters remain the same, only the coordinates are different. In case of lack of computational resources, we just submit a fresh job. The scripts to parse the outputs are packaged as docker images with defined input and output interfaces. The data-parsing workflow of Airavata has the capability to load those docker images and run containers using each job output file as input. After running the designated docker container, the parsed output for each child job is stored back in the gateway storage for future analysis.



\section{DISCUSSION AND CONCLUSION}
The initial testing of the GAMESS application and the pre- and post-parameter sweep implementations were done using a Jetstream virtual cluster\cite{coulter2019}, and construction of the final job scripts as required by the GAMESS application were configured. The GAMESS output from Stampede2 was validated against runs on the Jetstream cloud resource and also from a local computational cluster at the University of Michigan. During testing, it was determined that using the most recent version of GAMESS was causing inconsistencies, and GAMESS 2019 R02 was selected as the final version to use consistently. The GAMESS software was missing atomic radii and parameters for some of the solute atoms that were required for the calculation of dispersion and repulsion energies. Those atomic radii and parameters were supplied in the input explicitly. The computations took about 30 runs with 120 jobs each including some re-submissions to cover a total 5640 ligand and complex executions. A small number of jobs ($<$ 0.4\%) needed to be rerun as separate jobs due to convergence issues. 

Future directions for this project include:
\begin{enumerate}
    \item Computing the total binding energies and identifying drugs that are predicted to bind to the SARS-CoV-2 M$^{\rm pro}$ most favorably.
    \item Downstream analysis of the pair-wise interaction energies and its decomposition into energy components to identify key residues at the binding pocket and the chemical nature of interaction.
    \item Analyzing the atomic charges to explore ligand-specific polarization effects at the atomic and molecular level.
    \item Using the outcome from the above analyses at the lead optimization stage and also for generating new drug candidates targeting SARS-CoV-2 M$^{\rm pro}$.   
    \item Creating a searchable database that stores the key molecular data from GAMESS log files, e.g., atomic charges, pairwise energies, and solvation free energies. Access to this data will be important for training machine learning models capable of reproducing the key results from FMO calculations.
    \item Using the COVID-QM Gateway to apply FMO calculations to additional biomolecular systems using proteins, nucleic acids, and their complexes.
\end{enumerate}


\begin{acks}
This work used resources, services, and support provided via the COVID-19 HPC Consortium (https://covid19-hpc-consortium.org/).
\end{acks}

\bibliographystyle{ACM-Reference-Format}
\bibliography{bib-pearc21-fmo-covid19}

\end{document}